\documentclass[english,twoside,a4paper,10pt]{article}

\usepackage[latin1]{inputenc}
\usepackage[T1]{fontenc}
\usepackage{amsmath}
\usepackage{amsfonts}
\usepackage{graphicx}
\usepackage{subfigure}
\usepackage{a4wide}
\usepackage{amssymb}
\usepackage{fancyhdr}
\usepackage{mathrsfs}
\usepackage[toc,page]{appendix}

\begin{document}

\title{\bf Reconstruction of Inflation Models}
\author{ 
Ratbay Myrzakulov$^1$\footnote{Email: rmyrzakulov@gmail.com},\,\,\,
Lorenzo Sebastiani$^1$\footnote{E-mail address: l.sebastiani@science.unitn.it
},\,\,\,
Sergio Zerbini$^2$\footnote{E-mail address:zerbini@science.unitn.it
}\\
\\
\begin{small}
$1$ Department of General \& Theoretical Physics and Eurasian Center for
\end{small}\\
\begin{small} 
Theoretical Physics, Eurasian National University, Astana 010008, Kazakhstan
\end{small}\\
\begin{small}
$^2$ Dipartimento di Fisica, Universit\`a di Trento, Italy and
\end{small}\\
\begin{small}
  TIFPA, Istituto Nazionale di Fisica Nucleare, Trento, Italy
\end{small}\\
}

\date{}

\maketitle


\begin{abstract}
In this paper, we reconstruct viable inflationary models by starting from spectral index and tensor-to-scalar ratio from Planck observations. We analyze three different kinds of models: scalar field theories, fluid cosmology and $f(R)$-modified gravity. We recover the well known $R^2$-inflation in Jodan frame and Einstein frame representation, the massive scalar inflaton models and two models of inhomogeneous fluid. A model of $R^2$-correction to Einstein's gravity plus a "cosmological constant" with an exact solution for early time acceleration is reconstructed.
\end{abstract}



\tableofcontents
\section{Introduction}

It is well known that all the cosmological observations~\cite{WMAP, Planck} show with high accuracy that the universe is expanding in an accelerated way. Moreover, it is well accepted the idea according to which the universe, after the Big Bang, underwent an other period of strong accelerated expansion, namely the inflation (see Refs.~\cite{Mukbook,rev1, rev2} for review). 
The inflationary universe was first proposed by Guth and Sato
~\cite{Guth, Sato} to solve some problems related to the initial conditions
of Friedmann universe:
today, despite to the constraints that a viable theory must satisfy to fit the cosmological data,
the arena of the inflationary models is quite large.
In general, to reproduce the accelerated expansion of inflation, we may use a
scalar field representation, work with fluid cosmology, or modify the gravitational action of General Relativity introducing a repulsive force at high curvature.

In chaotic inflation~\cite{chaotic}
an homogeneous scalar field, namely the inflaton, is subjected to a potential and drives the accelerated expansion when its magnitude is negative and very large and satisfies some suitable conditions.
After this stage,
the field rolls down towards a potential minimum and inflation ends, such that the reheating process for the particle production starts supported by oscillations of the field
~\cite{buca1, buca2}. 

Fluid cosmology is the simplest approach to any desired cosmological scenario supported by non-perfect fluids different to standard matter and radiation~\cite{fluidOd}--\cite{Alessia(2)}: 
some applications to inflation can be found in Refs.~\cite{mioinfl1,mioinfl2}.

Finally, it is reasonable to expect that inflation 
might be  related to quantum corrections to Einstein's theory at high curvature. Thus, one may  take into account 
higher derivative terms of the curvature invariants (Ricci tensor, contractions of Ricci an Riemann tensor, Gauss-Bonnet, the ``square'' of the Weyl tensor...) in the gravitational action~\cite{q1, q1bis, q3}. 
We also mention that it is possible to study chaotic inflation in the framework of modified gravity (see Ref.~\cite{mioultimo}).
For reviews on modified gravity theories see Refs.~\cite{mod1, mod2, mod3}. 

In this work, following the approach presented in  Refs.~\cite{muk1, muk2}, we would like to start from the cosmological data and reconstruct the models able to reproduce a viable inflation in the different representations. In the specific, the last observations constrain the spectral index
as $n_s\sim 1-2/\mathcal N$, and the tensor-to-scalar ratio of cosmological perturbations as $r\sim 1/\mathcal N$ or, better, as $r\sim 1/\mathcal N^2$, where $\mathcal N$ is the total $e$-folds number of inflation and must be $55<\mathcal N<65$ to have a sufficient amount of inflation (for de Sitter space-time $n_s=1$ and we have an ethernal inflation with $\mathcal N\rightarrow\infty$). Thus, we can see which models satisfy these constraints by using a reconstruction  technique.

The paper is organized as follows. In Section {\bf 2}, we express all the characteristic quantities of inflation in terms of the $e$-folds number left to the end of inflation $N$, and we will write the spectral index and tensor-to-scalar ratio in terms of the total amount of inflation $\mathcal N$. The two viable choices illustrated above are analyzed in Section {\bf 3} and Section {\bf 4}, where two possible scenarios are investigated in the related subsections: scalar field representation and fluid cosmology. A specific section (Section {\bf 5}) is devoted to $f(R)$-modified gravity, for which the spectral index and the tensor-to-scalar ratio must be recalcolated. For every class of models we reconstruct the specific examples which generate viable inflation.
The results are discussed in Section {\bf 6}.

We use units of $k_{\mathrm{B}} = c = \hbar = 1$ and denote the
gravitational constant, $G_N$, by $\kappa^2\equiv 8 \pi G_{N}$, such that
$G_{N}=1/M_{\mathrm{Pl}}^2$, $M_{\mathrm{Pl}} =1.2 \times 10^{19}$ GeV being the Planck Mass.


\section{Inflation: general feature}

The inflation is a period of cosmological expansion where gravity acts as a repulsive force bringing to a strong accelerated expansion with a finite event horizon and explains the thermalization of observable universe. In order to preserve the important predictions of the Standard Model, inflation must take place in a very short period ($t\sim10^{-35}-10^{-36}$ sec.) after the Big Bang. The early-time acceleration may be described by a large variety of models, and in this section we would like to recall some basic features of it. 

The flat Friedmann-Robertson-Walker (FRW) metric for homogeneous and isotropic universe reads
\begin{equation}
ds^2=-dt^2+a(t)^2\left(dx^2+dy^2+dz^2\right)\,,\label{metric}
\end{equation}
where $a\equiv a(t)$ is the scale factor depending on cosmological time $t$. Thus, the Friedmann equations are
\begin{equation}
\frac{3H^2}{\kappa^2}=\rho_\text{eff}\,,\quad 
-\frac{(3H^2+2\dot H)}{\kappa^2}=p_\text{eff}\,,\label{EOMs}
\end{equation}
where the dot denotes the derivative with respect to the time, $H(t)\equiv H=\dot a/a$ is the Hubble parameter and $\rho_\text{eff}\,,p_\text{eff}$ are the effective energy density and pressure of the universe. They can be the result of some fluid contents, scalar field, and  also modifications to gravitational action in terms of higher derivative corrections to Einstein's gravity. For every case under investigation we will provide the corresponding Lagrangian to derive such quantities, but up to now we will keep this general framework. We may also introduce an effective Equation of State (EoS) as
\begin{equation}
p_\text{eff}=\omega_\text{eff}\rho_\text{eff}\,,\label{EoS}
\end{equation}
where $\omega_\text{eff}$ is an effective EoS parameter and in general it is a function of the Hubble parameter and its derivatives. 
From (\ref{EOMs}) we have the conservation law 
\begin{equation}
\dot\rho_\text{eff}+3H\rho_\text{eff}(1+\omega_\text{eff})=0\,.\label{conslaw}
\end{equation}
Inflation is described by a (quasi) de Sitter space-time, when the Hubble parameter is near to the Planck Mass. It means that at the beginning of the inflation
\begin{equation}
1+\omega_\text{eff}\simeq 0\,,\label{1om}
\end{equation}
but not vanishing in order to have an exit from inflation~\cite{muk1, muk2}. We  may also require the positivity of $(1+\omega)$: if $\omega$ pass throught minus one, it may be a final attractor of the system and inflation never ends; on the other side, acceleration vanishes when the Strong Energy Condition (SEC) is violated with $-1/3<\omega_\text{eff}$. To describe the evolution of (\ref{1om}),  we introduce the $e$-folds number left to the end of inflation,
\begin{equation}
N=\ln\left[\frac{a(t_\text{f})}{a(t)}\right]\,,
\end{equation}
where $a_\text{f}\equiv a(t_\text{f})$ is the scale factor at the end of inflation with $t_\text{f}$ the related time.
The inflation ends when $(1+\omega_\text{eff})\sim\mathcal O(1)$ and for this reason we may assume the following Ansatz~\cite{muk1},
\begin{equation}
1+\omega_\text{eff}\simeq\frac{\beta}{(N+1)^\alpha}\,,\quad 0<\alpha\,,\beta\,.\label{Anomega}
\end{equation}
Here, $\beta$ is a number on the order of the unit. 

In terms of the $e$-folds left to the end of inflation the conservation law (\ref{conslaw}) reads
\begin{equation}
-\frac{d\rho_\text{eff}}{dN}+3\rho_\text{eff}(1+\omega_\text{eff})=0\,,
\end{equation}
where we have taken into account that $d/dt=-H(t) d/d N$ and we have used (\ref{EoS}). As a consequence, from the Ansatz (\ref{Anomega}) we find the following solutions
\begin{equation}
\rho_\text{eff}\simeq\rho_0(N+1)^{3\beta}\,,\quad\alpha=1\,,\label{rho1}
\end{equation}
\begin{equation}
\rho_\text{eff}\simeq\rho_0\exp\left[-\frac{3\beta}{(\alpha-1)(N+1)^{\alpha-1}}\right]\,,\label{rho2}
\quad\alpha\neq 1\,,
\end{equation}
with $\rho_0$ an integration constant corresponding to the effective energy density at the end of inflation at $N=0$ in the case of $\alpha=0$, and to the effective energy density at the beginning of inflation at $1\ll N$ in the case of $1<\alpha$. 
The quasi de Sitter solution of inflation evolves (at the first order approximation) with the Hubble flow functions
\begin{equation}
\epsilon_1=-\frac{\dot H}{H^2}\,,\quad 
\epsilon_2=-\frac{2\dot H}{H^2}+\frac{\ddot H}{H\dot H}\equiv \frac{\dot\epsilon_1}{H\epsilon_1}\,.\label{Hflow}
\end{equation}
First let us assume to deal with  situations where the so called Jordan farme coincides with the Einstein frame. Thus for the moment, we shall not consider
modified gravity models. Inflation takes place as soon as the quantities above remain very small (slow-roll approximation). In the specific must be,
\begin{equation}
\epsilon_1\ll 1\,,\quad |\epsilon_2|\ll 1\,,\label{slowrollregime}
\end{equation}
and acceleration ends when $\epsilon_1$ is on the order of the unit. By using (\ref{EOMs}) --(\ref{conslaw}) we find
\begin{equation}
\epsilon_1=\frac{3(1+\omega_\text{eff})}{2}\,,\quad
\epsilon_2=-\frac{d\ln [1+\omega_\text{eff}]}{d N}\,,
\end{equation}
or, by using Ansatz (\ref{Anomega}),
\begin{equation}
\epsilon_1\simeq\frac{3\beta}{2(N+1)^\alpha}\,,\quad
\epsilon_2\simeq\frac{\alpha}{N+1}\,.\label{flowflow}
\end{equation}
We see that if $1<\alpha$ the $\epsilon_1$ parameter is much smaller with respect to the $\epsilon_2$ parameter. This behaviour can be found in all the models where inflation can be treated at the perturbative level such that $\dot H^2\ll H |\ddot H|$, but in general the slow-roll conditions do not imply it. In principle, also the case $0<\alpha<1$ where $\epsilon_1$ is larger than $\epsilon_2$ is not excluded.

In order to solve the problem of initial conditions of the Friedmann
universe, it is necessary to have $\dot a_\mathrm{i}/\dot a_0<
10^{-5}$, where $\dot a_\mathrm{i}\,,\dot a_0$ are the time derivatives of the
scale factor at the Big Bang and today, respectively, and the anisotropy in our universe is on the order of $10^{-5}$. 
Since in the
decelerating universe $\dot a(t)$ decreases by a factor $10^{28}$, one has $\dot a_\mathrm{i}/\dot a_\mathrm{f}<10^{-33}$ and if inflation is
governed by a (quasi) de Sitter solution the number of $e$-folds at the beginning of inflation,
\begin{equation}
\mathcal N\equiv N |_{t=t_\text{i}}=\ln \left(\frac{a_\mathrm{f}(t_\text{f})}{a_\mathrm{i} (t_\text{i})}\right)\equiv\int^{t_\text{f}}_{t_\text{i}} H(t)
dt\,,\label{Nfolds}
\end{equation}
must be at least $\mathcal N\simeq 76$, but the spectrum of fluctuations of CMB say that it is enough $\mathcal N\simeq 55$ to have thermalization of observable universe. Thypically, it is required $55<\mathcal N<65$. 

Thus, at the end of inflation, the amplitude of the power spectrum of Newtonian potential (namely the fluctuations of the effective energy density) is given by
\begin{equation}
\Delta_{\mathcal R}^2=\frac{\kappa^2 H^2}{8\pi^2\epsilon_1}|_{N=\mathcal N}\equiv\frac{16}{9}G_N^2\left(\frac{\rho_\text{eff}}{(1+\omega_\text{eff})}\right)|_{N=\mathcal N}\,,\label{spectrum}
\end{equation}
where we have reintroduced the Newton Constant $G_N$. For the Ansatz (\ref{Anomega}) 
with (\ref{rho1})--(\ref{rho2})
one derives
\begin{equation}
\Delta_{\mathcal R}^2\simeq\frac{16}{9}G_N^2\frac{\rho_0(\mathcal N+1)^{3\beta+1}}{\beta}\,,\quad\alpha=1\,,\label{pow1}
\end{equation}
\begin{equation}
\Delta_{\mathcal R}^2\simeq\frac{16}{9}G_N^2\frac{\rho_0(\mathcal N+1)^{\alpha}\exp\left[-\frac{3\beta}{(\alpha-1)(\mathcal N+1)^{\alpha-1}}\right]}{\beta}\,,\quad\alpha\neq 1\,.\label{pow2}
\end{equation}
Since it must be $\Delta_{\mathcal R}^2\simeq 10^{-9}$, given the $e$-folds number $\mathcal N$, we may use the power spectrum to recover the effective energy density of the universe at the end of inflation, which reads as $\rho=\rho_0$ in the case of $\alpha=1$ and $\rho=\rho_0\exp\left[-3\beta/(\alpha-1)\right]$ in the case of $\alpha\neq 1$.

The spectral index $n_s$ and the tensor-to-scalar 
ratio $r$ are given by (at the first order),
\begin{equation}
n_s=1-2\epsilon_1|_{N=\mathcal N}-\epsilon_2|_{N=\mathcal N}
\equiv1-3(1+\omega_\text{eff})+\frac{d}{dN}\ln\left(1+\omega_\text{eff}\right)
\,,\quad r=16\epsilon_1|_{N=\mathcal N}
\equiv 24(1+\omega_\text{eff})
\,.\label{index}
\end{equation}
By using (\ref{Anomega}), one has~\cite{muk1},
\begin{equation}
n_s=1-\left[\frac{3\beta+\alpha(\mathcal N+1)^{\alpha-1}}{(\mathcal N+1)^{\alpha}}\right]\,,
\quad r=\frac{24\beta}{(\mathcal N+1)^{\alpha}}\,.
\end{equation}
In the plane $(n_s,r)$, one has the curve defined by
\begin{equation}
r=24 \beta \left(1-n_s -\frac{r}{8}\right)^{\alpha}\,.
\end{equation}
Note that the spectral index $n_s$ is smaller than one, since the Hubble flow functions are both positive. This is always true if we want a graceful exit from inflation. 

The last observations by the Planck satellite~\cite{Planck} constrain the spectral index and the tensor-to-scalar ratio as
$n_{\mathrm{s}} = 0.9603 \pm 0.0073\, (68\%\,\mathrm{CL})$ and 
$r < 0.11\, (95\%\,\mathrm{CL})$. It means
\begin{equation}
0.0324<\left[\frac{3\beta+\alpha(\mathcal N+1)^{\alpha-1}}{(\mathcal N+1)^{\alpha}}\right]<0.0470\,,\quad \frac{24\beta}{(\mathcal N+1)^{\alpha}}<0.11\,.
\label{Presults}
\end{equation}
We have the following cases
\begin{equation}
1-n_s\simeq\frac{3\beta+1}{(\mathcal N+1)}\,,\quad \alpha=1\,,\label{1}
\end{equation}
\begin{equation}
1-n_s\simeq\frac{\alpha}{(\mathcal N+1)}\,,\quad1<\alpha\,,\label{2}
\end{equation}
\begin{equation}
1-n_s\simeq\frac{3\beta}{(\mathcal N+1)^\alpha}\,,\quad\alpha<1\,.\label{3}
\end{equation}
Therefore, the cases $\alpha=1$ and $\beta=1/3$ and $\alpha=2$ with $\mathcal N\simeq 60$ are viable~\cite{muk2}, the second one with a tensor-to-scalar ratio much smaller. In the first case, since $r=4(1-n_s)$, the tensor-to-scalar ratio is slightly larger than the Planck results when the spectral index is in agreement with them, but, since the correct value of this parameter is still a debated question, in our work we will take in consideration also the class of models which realizes such a configuration.
If $\alpha<1$, the tensor-to-scalar ratio is too big: for example, if $\alpha=3/4$ and $\beta=1/3$ with $\mathcal N\simeq 60$, the spectral index staisfies the Planck contraints, but the tensor-to-scalar ratio is $r=0.37$.

Let us analyze now how different models can be reconstructed to reproduce a viable inflationary scenario. 

\section{Inflation with spectral index $1-n_s=(3\beta+1)/(\mathcal N+1)$: case $\alpha=1$}

In this Section, we would like to analyze different models to realize viable inflation with spectral index (\ref{1}): it means that the Hubble flow functions (\ref{Hflow}) are on the same order of magnitude and the EoS parameter in (\ref{EoS}) can be written as
\begin{equation}
\omega_\text{eff}=-1+\frac{1}{3}\left(\frac{\rho_0}{\rho_\text{eff}}\right)\,,\label{omegaalpha1}
\end{equation}
where we have used (\ref{Anomega}) and (\ref{rho1}) with $\alpha=1\,,\beta=1/3$ to satisfy the Planck results. Here, $\rho_0$ is the effective energy density at the end of inflation and when $\rho_\text{eff}=\rho_0$ acceleration ends.
The Hubble flow functions (\ref{flowflow}) read in this case
\begin{equation}
\epsilon_1=\frac{1}{2(N+1)}\,,\quad\epsilon_2=\frac{1}{N+1}\,,\label{flow1}
\end{equation}
and the spectral index and the tensor-to scalar ratio are given by
\begin{equation}
n_s=1-\frac{2}{\mathcal N+1}\,,\quad r=\frac{8}{\mathcal N+1}=4(1-n_s)\,.\label{spec1}
\end{equation}
First of all, we will revisit chaotic inflation in scalar field theories.

\subsection{Inflation with  scalar field\label{chaotic}}

Inflation may be  realized by the inclusion of a  minimally coupled scalar field $\phi$, dubbed inflaton, subjected to potential $V(\phi)$ whose general Lagrangian reads
\begin{equation}
\mathcal L_\phi=-\frac{1}{2}g^{\mu\nu}\partial_{\mu}\phi\partial_{\nu}\phi-V(\phi)\,.
\end{equation}
On flat FRW metric the Equations of motion (EOMs) are given by (\ref{EOMs}) with the following identification
\begin{equation}
\rho_\text{eff}=\frac{\dot\phi^2}{2}+V(\phi)\,,\quad p_\text{eff}=\frac{\dot\phi^2}{2}-V(\phi)\,,\quad\omega_\text{eff}=\frac{\dot\phi^2-2V(\phi)}{\dot\phi^2+2V(\phi)}\,.\label{EOMsb}
\end{equation}
We get from the first equation in (\ref{EOMs}) and from the conservation law (\ref{conslaw})
\begin{equation}
\frac{3H^2}{\kappa^2}=\frac{\dot\phi^2}{2}+V(\phi)\,,\quad
\ddot\phi+3H\dot\phi=-V'(\phi)\,,\label{EOMsfield}
\end{equation}
where the prime denotes the derivative respect to $\phi$. Chaotic inflation is realized for negative and arbitrary large values of the field when the slow-roll approximation is valid. For scalar field representation one introduces the slow-roll parameters
\begin{equation}
\epsilon=-\frac{\dot H}{H^2}\,,\quad\eta=-\frac{\dot H}{H^2}-\frac{\ddot H}{2 H\dot
H}\equiv2\epsilon-\frac{1}{2\epsilon H}\dot\epsilon\,,\label{slowrollpar}
\end{equation}
which are related to the Hubble flow functions (\ref{Hflow}) as
\begin{equation}
\epsilon=\epsilon_1\,,\quad \eta=-\frac{\epsilon_2}{2}+2\epsilon_1\,.\label{slowrollpar2}
\end{equation}
Thus, in the slow-roll regime (\ref{slowrollregime}) we get
\begin{equation}
\epsilon\ll 1\,,\quad |\eta|\ll 1\,.\label{s2}
\end{equation}
To realize the quasi de Sitter solution of inflation with $\omega_\text{eff}\simeq -1$, the kinetic energy of the field must be much smaller respect to the potential,
\begin{equation}
\dot\phi^2\ll V(\phi)\,,\quad |\ddot\phi|\ll 3H \dot\phi\,,\label{phiV}
\end{equation}
such that equations (\ref{EOMsfield}) read
\begin{equation}
\frac{3 H^2}{\kappa^2}\simeq V(\phi)\,,\quad 3H\dot\phi\simeq-V'(\phi)\,,
\label{EOMsfield2}
\end{equation}
and the slow-roll parameters (\ref{slowrollpar}) and the $e$-folds $\mathcal N$ in (\ref{Nfolds}) can be derived as
\begin{equation}
\epsilon=\frac{1}{2\kappa^2}\left(\frac{V'(\phi)}{V(\phi)}\right)^2\,,\quad 
\eta=\frac{1}{\kappa^2}\left(\frac{V''(\phi)}{V(\phi)}\right)\,,\quad
\mathcal N=\kappa^2\int^{\phi_i}_{\phi_e}\frac{V(\phi)}{V'(\phi)}d\phi\,.\label{ciao}
\end{equation}
Finally, the spectral index and the tensor-to-scalar ratio in (\ref{index}) are given by
\begin{equation}
n_s=1-6\epsilon|_{N=\mathcal N}+2\eta|_{N=\mathcal N}\,,\quad r=16\epsilon|_{N=\mathcal N}\,.\label{ind1}
\end{equation}
For the case under investigation (\ref{1}) we have
\begin{equation}
\epsilon=\eta=\frac{1}{2(N+1)}\,,\quad n_s=1-\frac{2}{\mathcal N+1}\,,\quad r=\frac{8}{\mathcal N+1}=4(1-n_s)\,,
\label{spectral1}
\end{equation}
where the spectral index and the tensor-to-scalar ratio are obviously the same of (\ref{spec1}).
For large e-folds $N$ during inflation the slow-roll parameters are very small and for $\mathcal N\simeq 60$, as we have seen, the spectral index is in good agreement, but  the tensor-to-scalar ratio seems slightly out.
Inflation ends when the slow-roll conditions in (\ref{s2}) are not more valid. Typically, the field slowly moves toward a potential minimum and starts to oscillate beginning the rehating process for particle production.\\
\\
Reconstructing a model for chaotic inflation which realizes (\ref{1}) with $\beta=1/3$ and therefore the spectral index and the tensor-to -scalar ratio in (\ref{spectral1}) is quite simple. From (\ref{omegaalpha1}) and the last expression in (\ref{EOMsb}) we get
\begin{equation}
\phi=\phi_\text{i}+\sqrt{\frac{\rho_0}{3}}(t-t_\text{i})\,.\label{ex0}
\end{equation}
Here, the integration constant $\phi_\text{i}$ corresponds to the (negative) value of the field at the beginning of inflation when $t=t_\text{i}$. Thus, in the slow roll approximation (\ref{phiV}), we can use (\ref{EOMsfield2}) to find
\begin{equation}
V(\phi)=\frac{(-\phi)^2 m^2}{2}\,,\quad H\simeq\frac{m(-\phi)\kappa}{\sqrt{6}}\,,
\end{equation} 
where we have introduced the constant term
\begin{equation}
m^2=\frac{\kappa^2\rho_0}{2}\,.\label{m2}
\end{equation}
This theory corresponds to a quadratic scalar field inflation. The boundary value of the field defines the total amount of inflation, namely $\mathcal N$ throught relation (\ref{rho1}) with $\beta=1/3$ as
\begin{equation}
\rho_i\simeq\frac{2m^2}{\kappa^2}(\mathcal N+1)\,,
\end{equation}
where we have used (\ref{m2}). 
When $\rho_\text{eff}=\rho_0\equiv2m^2/\kappa^2$, the effective EoS parameter $\omega_\text{eff}=-1/3$ and inflation ends: at this point, the potential $V(\phi)=\rho_0/3\equiv 2m^2/(3\kappa^2)$ is equal to $\dot\phi^2$ and the slow-roll approximation (\ref{phiV}) is not more valid. Finally, the value of $m^2$ and therefore of $\rho_0$ must be consistent with the amplitude of the power spectrum in (\ref{pow1}) as $\Delta_{\mathcal R}\simeq 10^{-9}$.\\
\\
The quadratic potential is not the only one which realizes chaotic inflation with $\alpha=1$ in (\ref{1}). If we relax the condition on $\beta$ in (\ref{omegaalpha1}) such that
\begin{equation}
\omega_\text{eff}=-1+\beta\left(\frac{\rho_0}{\rho_\text{eff}}\right)^{\frac{1}{3\beta}}\,,\label{omegaalpha1bis}
\end{equation}
where $\beta\neq 1/3$ remains a number on the order of the unit, in the slow-roll approximation (\ref{phiV}) we find the following relation between the kinetic energy of the field and the potential,
\begin{equation}
\dot\phi\simeq\frac{\sqrt{\beta}\rho_0^{1/(6\beta)}}{V(\phi)^{\frac{1-3\beta}{6\beta}}}\,.
\end{equation}
For $\beta=1/3$ one derives (\ref{ex0}) again.
Thus, by using the second equation in (\ref{EOMsfield2}) we obtain
\begin{equation}
V(\phi)=\frac{(-\phi)^{6\beta}(3\kappa^2)^{3\beta}\rho_0}{6^{6\beta}\beta^{3\beta}}\,,
\end{equation}
namely a power-law potential. For example, for $\beta=2/3$ we get the quartic potential
\begin{equation}
V=\frac{\lambda(-\phi)^4}{4}\,,\quad\lambda=\frac{\kappa^4\rho_0}{16}\,,
\end{equation} 
with the slow-roll parameters (\ref{slowrollpar})--(\ref{slowrollpar2}) and spectral index and tensor-to-scalar-ratio (\ref{ind1})
\begin{equation}
\epsilon=\frac{1}{N+1}\,,\quad\eta=\frac{3}{2(N+1)}\,,\quad
n_s=1-\frac{3}{\mathcal N+1}\,,
\quad r=\frac{16}{\mathcal N+1}=\frac{16}{3(1-n_s)}\,.\label{ex1}
\end{equation}
Also in this example, for large e-folds $N$ during inflation the slow-roll parameters are very small and for $\mathcal N\simeq 60$ the spectral index satisfies the Planck data, but the tensor-to-scalar ratio is too large. All the potentials based on power-law function belong to the class (\ref{1}), according with Ref.~\cite{muk1}. The only potential which is not disfavoured by Plank data seems to be the linear potential, which is realized by the choice 
$\beta=\frac{1}{6}$. In this case , one also has $r=\frac{8}{3}(1-n_s)$.

\subsection{Fluid cosmology}

The simplest description of inflation is provided by exotic fluid with Equation of State different to the one of ordinary matter and radiation~\cite{mioinfl1,mioinfl2}. Since the inflation is described by a quasi de-Sitter solution and inflation must ends, the EoS parameter of such a fluid cannot be a constant (like for perfect fluids) and must depend on the energy density. If we identify $\rho_\text{eff}$ and $p_\text{eff}$ with the energy density $\rho$ and the pressure $p$ of the fluid, the Equation of State (\ref{EoS}) with (\ref{omegaalpha1}) read
\begin{equation}
p=\omega(\rho)\rho\,,\quad\omega(\rho)=-1+\frac{1}{3}\left(\frac{\rho_0}{\rho}\right)\,,
\end{equation}
namely the EoS parameter $\omega(\rho)$ of the (inhomogeneous) fluid depends on the energy density. Thus, from (\ref{EOMs}) and (\ref{conslaw}), one has
\begin{equation}
H=\sqrt{\frac{\kappa^2}{3}}\left[h_0-\frac{\rho_0}{2}\sqrt{\frac{\kappa^2}{3}}(t-t_\text{i})\right]\,,\quad
\rho=\left[h_0-\frac{\rho_0}{2}\sqrt{\frac{\kappa^2}{3}}(t-t_\text{i})\right]^2\,.
\end{equation}
Inflation takes place when $t$ is close to $t_\text{i}$ and $H=\sqrt{\kappa^2/3} h_0$, where $h_0$ is the constant value of the de Sitter solution of early-time acceleration. The energy density of the fluid and therefore the Hubble parameter decrease. The Hubble flow functions (\ref{Hflow}) read
\begin{equation}
\epsilon_1=\frac{\rho_0}{2\rho}\,,\quad\epsilon_2=\frac{\rho_0}{\rho}\,.
\end{equation}
Since the scale factor can be written as
\begin{equation}
a(t)=a_\text{f}\exp\left[1-\frac{\rho}{\rho_0}\right]\,,
\end{equation}
where $a_\text{f}$ is the scale factor at the end of inflation when $\rho=\rho_0$, we see that
\begin{equation}
N+1=\frac{\rho}{\rho_0}\,,
\end{equation}
and we recover (\ref{flow1})--(\ref{spec1}). 

Also in this case, other inhomogeneous fluid models reproducing inflation with Hubble flow functions on the same order can be found by relaxing the condition on $\beta$,
namely $\beta\neq 1/3$ in (\ref{omegaalpha1}) like in (\ref{omegaalpha1bis}),
\begin{equation}
\omega=-1+\beta\left(\frac{\rho_0}{\rho}\right)^{\frac{1}{3\beta}}\,,\label{omegaalpha1bisbis}
\end{equation}
which seems to define an extended Chaplygin gas~\cite{C1,C2}.
In such a case we derive
\begin{equation}
H=\sqrt{\frac{\rho_\text{i}\kappa^2}{3}}\exp\left[-\sqrt{\frac{\rho_0\kappa^2}{3}}(t-t_\text{i})\right]
\,,\quad
\rho=\rho_\text{i}\exp\left[-2\sqrt{\frac{\rho_0\kappa^2}{3}}(t-t_\text{i})\right]\,,
\quad\beta=\frac{2}{3}\,,
\end{equation}
\begin{eqnarray}
H&=&\sqrt{\frac{\kappa^2}{3}}\left[\tilde h_0-\left(\frac{2-3\beta}{2}\right)\rho_0^{\frac{1}{3\beta}}\sqrt{\frac{\kappa^2}{3}}\left(t-t_\text{i}\right)\right]^{\frac{3\beta}{2-3\beta}}
\,,\nonumber\\
\rho&=&\left[\tilde h_0-\left(\frac{2-3\beta}{2}\right)\rho_0^{\frac{1}{3\beta}}\sqrt{\frac{\kappa^2}{3}}\left(t-t_\text{i}\right)\right]^{\frac{6\beta}{2-3\beta}}\,,\quad\beta\neq\frac{2}{3}\,,
\end{eqnarray} 
where $\rho_\text{i}$ (in this case, the energy density of the fluid at the beginning of inflation) and $\tilde h_0$ are integration constants. 
For example, for the case $\beta=2/3$ above, one recovers the spectral index and the tensor-to-scalar ratio in (\ref{ex1}).

\section{Inflation with spectral index $1-n_s=2/(\mathcal N+1)$: case $\alpha=2$}

Now we would like to analyze different models to realize inflation with spectral index (\ref{2}), where the Hubble flow function $\epsilon_1$ in (\ref{Hflow}) is much smaller than $\epsilon_2$. Since the tensor-to-scalar ratio is proportional to $r\sim1/(\mathcal N+1)^2$, every value of $\beta$ on the order of the unit satisfies the Planck data. In this case, the EoS parameter in (\ref{EoS}) can be written as
\begin{equation}
\omega_\text{eff}=-1+\frac{1}{9\beta}\log\left[\frac{\rho_\text{eff}}{\rho_0}\right]^2\,,\label{omegaalpha2}
\end{equation}
where we have used (\ref{Anomega}) and (\ref{rho1}) with $\alpha=2$. Now $\rho_0$ is the effective energy density at the beginning of inflation, when $\omega=-1$.
The Hubble flow functions (\ref{flowflow}) read 
\begin{equation}
\epsilon_1=\frac{3\beta}{2(N+1)^2}\,,\quad\epsilon_2=\frac{2}{N+1}\,,\label{flow2}
\end{equation}
and the spectral index and the tensor-to scalar ratio are derived as
\begin{equation}
n_s=1-\frac{2}{\mathcal N+1}\,,\quad r=\frac{24\beta}{\left(\mathcal N+1\right)^2}=6\beta(1-n_s)^2\,.\label{spec2}
\end{equation}
Let us see some applications.

\subsection{Scalar  inflation}

Let us come back to scalar field inflation described in \S~\ref{chaotic}. From (\ref{omegaalpha2}) and the last expression in (\ref{EOMs}), if we use the slow-roll approximation (\ref{phiV}), we have
\begin{equation}
\dot\phi\simeq\frac{\sqrt{V(\phi)}}{3\sqrt{\beta}}\log\left[\frac{\rho_0}{V(\phi)}\right]\simeq
\frac{\sqrt{V(\phi)}}{3\sqrt{\beta}}\left(\frac{\rho_0}{V(\phi)}-1\right)\,,
\end{equation}
where we have used the fact that during inflation $V(\phi)$ is close to $\rho_0^-$.
Thus, from the second equation in (\ref{EOMsfield2}) we get
\begin{equation}
V(\phi)=\rho_0\left(1-c_1 \text{e}^{\sqrt{\kappa^2/(3\beta)}\phi}
\right)\,,
\end{equation}
where $c_1$ is a constant. In the case $c_1=2$ and $\beta=1/2$ we recover inflation in the scalar field Einstein frame representation of Starobinsky model and we obtain for the slow-roll parameters introduced in (\ref{slowrollpar}),
\begin{equation}
\epsilon=\frac{3}{4(N+1)^2}\,,\quad\eta\simeq-\frac{1}{(N+1)^2}\,.
\end{equation}
The spectral index and the tensor-to-scalar ratio read
\begin{equation}
n_s=1-\frac{2}{\mathcal N+1}\,,\quad r=\frac{12}{\left(\mathcal N+1\right)^2}=3(1-n_s)^2\,,
\label{spectralstaro}
\end{equation}
according with the results of Starobinsky inflation~\cite{Staro} in the Einstein frame, and we will see in later  also in Jordan frame. 
Furthermore, in Ref.~\cite{mioStaro} has been shown that also the more general class of $R^2$-models with cosmological constant reproduces the inflation in the same way as the Starobinsky model and the cosmological constant does not play here any role, since it produces a (negligible) term proportional to $\sim\exp\left[2\sqrt{\kappa^2/(3\beta)}\phi\right]$ in the potential.

\subsection{Fluid cosmology}

The Equation of State (\ref{EoS}) with (\ref{omegaalpha2}) assumes the following form for inhomogneous fluid,
\begin{equation}
p=\omega(\rho)\rho\,,\quad\omega(\rho)=-1+\frac{1}{9\beta}\log\left[\frac{\rho}{\rho_0}\right]^2
\simeq -1+\frac{1}{9\beta}\left[\left(\frac{\rho_0}{\rho}\right)-1\right]^2\,.
\end{equation}
The asymptotic solution of (\ref{EOMs}) and (\ref{conslaw}) read
\begin{equation}
H=\sqrt{\frac{\kappa^2}{3}}\sqrt{\left(
1-
\frac{3\sqrt{3}\beta}{\sqrt{\kappa^2\rho_0}(t_\text{e}-t)}\right)}\,,\quad
\rho\simeq\rho_0\left(
1-
\frac{3\sqrt{3}\beta}{\sqrt{\kappa^2\rho_0}(t_\text{e}-t)}\right)\,,
\end{equation}
in the limit $t\ll t_\text{e}$, where $t_\text{e}$ is the time at the end of inflation (in fact, its duration).
Thus, the Hubble flow paramters (\ref{Hflow}) are derived as
\begin{equation}
\epsilon_1\simeq\frac{9\beta}{2(t_\text{e}-t)^2\kappa^2\rho_0}\,,
\quad
\epsilon_2\simeq\frac{2\sqrt{3}}{\sqrt{\kappa^2\rho_0}(t_\text{e}-t)}\,,
\end{equation}
and the spectral index and the tensor-to-scalar ratio (\ref{index}) finally are given by
\begin{equation}
n_s\simeq 1-\frac{2\sqrt{3}}{(t_\text{e}-t)\sqrt{\kappa^2\rho_0}}\,,
\quad
r\simeq\frac{72\beta}{(t_\text{e}-t)^2\kappa^2\rho_0}\,.
\end{equation}
By taking into account that
\begin{equation}
a(t)\simeq a_\text{f}\exp\left[\sqrt{\frac{\kappa^2\rho_0}{3}}(t-t_\text{e})\right]\,,\quad 
N\simeq\sqrt{\frac{\kappa^2\rho_0}{3}}(t_\text{e}-t)\,,
\end{equation}
one recovers
(\ref{spec2}).

\section{$f(R)$-modified gravity}

An alternative description of the early-time acceleration is given by the modified theories of gravity, where an arbitrary function of some curvature invariant is added to the Hilbert-Einstein term in the action of General Relativity. This kind of corrections may arise from quantum effects or be inspired by string-theories and it is expected that plays a fundamental role at high curvatures, during the inflation. Here, we would like to analyze the simplest class of such models, where the modification depends on the Ricci scalar only. 

Let us consider the following mofified gravitational Lagrangian,
\begin{equation}
\mathcal L_{grav}=\frac{R^2}{2\kappa^2}+f(R)\,,
\end{equation}
where $f(R)$ is a function of the Ricci scalar $R$. 
On FRW metric (\ref{metric}) the modification to gravity can be encoded in the effective energy density and pressure which appear in (\ref{EOMs}) by making the following identification
\begin{eqnarray}
\rho_{\mathrm{eff}} &=& \biggl[(Rf_{R}
-f)-6H\dot{f}_{R}
-6H^{2}f_R
\biggr]\,,\label{rhoMG}\\
p_{\mathrm{eff}} &=& \biggl[
(f-R f_{R})+4H\dot{f}_{R}+2\ddot{f}_{R}
+(4\dot{H}+6H^{2})f_R                      
\biggr]\,.\label{pMG}
\end{eqnarray}
Here, $f(R)\equiv f$ and the subscript `$R$' is the derivative with respect to the Ricci scalar. 

As well known, for a modified theory of gravity, the Einstein frame does not coincides with the Jordan frame, where the theory is defined. Thus,  the Hubble flow functions (\ref{Hflow}) have to be  replaced by the variables \cite{corea} (for a review of inflation in the framework of $f(R)$-gravity see the Ref.~\cite{DeFelice}),
\begin{equation}
\epsilon_1=-\frac{\dot H}{H^2}\,,\quad\epsilon_3=\frac{\kappa^2\dot f_R}{H\left(1+2\kappa^2f_R\right)}\,,\quad
\epsilon_4=\frac{\ddot f_R}{H\dot f_R}\,,
\end{equation}
and the spectral index and the tensor-to-scalar ratio are given by
\begin{equation}
n_s=1-4\epsilon_1+2\epsilon_3-2\epsilon_4\,,\quad r=48\epsilon_3^2\,.
\end{equation}
From the EOMs (\ref{EOMs}) with (\ref{rhoMG})--(\ref{pMG}) one has
\begin{equation}
\epsilon_1=-\epsilon_3(1-\epsilon_4)\,.
\end{equation}
During inflation $|\epsilon_{1,3,4}|\ll 1$ and $\epsilon_1\simeq-\epsilon_3$, such that $\epsilon_4\simeq-3\epsilon_1+\dot\epsilon_1/(H\epsilon_1)$.
It follows
\begin{equation}
n_s=1-2\epsilon_2\,,\quad r=48\epsilon_1^2\,,
\end{equation}
where we have reintroduced $\epsilon_2$ in (\ref{Hflow}). Thus, in analogy with (\ref{index}), we find
\begin{equation}
n_s=1-2\epsilon_2|_{N=\mathcal N}
\equiv1-2\frac{d}{dN}\ln\left(1+\omega_\text{eff}\right)
\,,\quad r=48\epsilon_1^2|_{N=\mathcal N}
\equiv 108(1+\omega_\text{eff})^2\,.
\end{equation}
By using (\ref{Anomega}), we derive
\begin{equation}
n_s=1-\frac{2\alpha}{(\mathcal N+1)}\,,
\quad r=\frac{108\beta^2}{(\mathcal N+1)^{2\alpha}}\,,
\end{equation}
and the choice $\alpha=2$ with $\mathcal N\simeq 60$ staisfies the Planck results in (\ref{Presults}). The corresponding $\omega_\text{eff}\equiv p_\text{eff}/\rho_\text{eff}$ parameter is given by $\omega_\text{eff}=-1+\beta\left(\rho_0/\rho_\text{eff}\right)^{1/(3\beta)}$, but for simplicity we will set $\beta=1/3$ recovering $\omega_\text{eff}=-1+\rho_0/(3\rho_\text{eff})$
as in (\ref{omegaalpha1}). It follows from (\ref{rhoMG})--(\ref{pMG}),
\begin{equation}
4\dot H f_R-2H\dot f_R+2\ddot f_R=\frac{\rho_0}{3}\,,
\label{omcondm}
\end{equation}
with
\begin{equation}
R=12H^2+6\dot H\,.
\end{equation}
If we use (\ref{rho1}) or (\ref{rho2}) in the first equation of (\ref{EOMs}), we can express the Hubble parameter and its time derivative in terms of the e-folds left to the end of inflation $N$. In the case of (\ref{rho1}) with $\beta=1/3$ one has
\begin{equation}
H=\sqrt{\frac{\kappa^2\rho_0}{3}}\sqrt{N+1}\,,\quad\dot H=-\frac{\kappa^2\rho_0}{6}\,.\label{solpart}
\end{equation}
Thus, we can reconstruct the modified gravity models which realize such a configuration and we get from (\ref{omcondm}),
\begin{equation}
-f_R+\left(\frac{2N+3}{2}\right)\frac{d f_R}{d N}+(N+1)\frac{d^2 f_R}{d N^2}=\frac{1}{2\kappa^2}\,,\quad f_R=-\frac{1}{2\kappa^2}+\gamma\left(\frac{3}{2}+N\right)\,,
\end{equation}
$\gamma$ being an integration constant.
Since $R=\kappa^2\rho_0(4N+3)$, one has
\begin{equation}
f_R(R)=\frac{R\gamma-2\rho_0+3\gamma\kappa^2\rho_0}{4\kappa^2\rho_0}\,,\quad
f(R)=\frac{3R\gamma}{4}-\frac{R}{2\kappa^2}+\frac{R^2\gamma}{8\kappa^2\rho_0}
+\lambda\,,\label{eqexf}
\end{equation}
where $\lambda$ is a ``cosmological constant''. From the last expression we obtain for the effective energy density (\ref{rhoMG}) respect to (\ref{solpart}),
\begin{equation}
\rho_\text{eff}=-\lambda+\rho_0(1+N)+\frac{\gamma\kappa^2\rho_0}{8}\,,\label{rholast}
\end{equation}
such that it has to be
\begin{equation}
\lambda=\frac{\gamma\kappa^2\rho_0}{8}\,,\label{cc}
\end{equation}
in order to satisfy the first equation in (\ref{EOMs}). Finally, if we set $\gamma=2/(3\kappa^2)$, we obtain
\begin{equation}
f(R)=\frac{R^2}{12\kappa^4\rho_0}+\frac{\rho_0}{12}\,,\label{mnew}
\end{equation}
corresponding to an extension of the so called ``Starobinski model'' where $f(R)=\tilde\gamma R^2$, being $\tilde\gamma$ a constant~\cite{Staro}. For such a model, we recover the spectral index and the tensor-to-scalar ratio in (\ref{spectralstaro}). We also note that, in the limit $1\ll N$, expression (\ref{rholast}) solves asymptotically the first Friedmann equation in (\ref{EOMs}) independently on $\lambda$ if
\begin{equation}
\gamma\kappa^2\ll 8(1+N)\,,
\end{equation}
and, in the case of $\gamma=2/(3\kappa^2)$, this is always true if $\kappa^2\rho_0\ll R$ or, since $\rho_0\sim1/\kappa^4$, $M_{Pl}^2/8\pi=1/\kappa^2\ll R$. In the Starobisky-like models the Planck Mass in $\kappa^2$ is replaced by a more general mass term $M^2$, and during inflation the term $R^2$ can be considered much bigger than the Hilbert-Einstein contribute to the action. 
As it has been found in Ref.~\cite{mioStaro}, this class of models produces inflation in slow-roll approximation in the same way as the Starobinsky model. Here, we see also that, if we add a cosmological constant like in (\ref{cc}), we have an exact accelerated solution for inflation. In terms of the cosmological time, this exact solution reads
\begin{equation}
H(t)=H_0-\frac{\kappa^2\rho_0}{6}(t-t_\text{i})\, \quad a(t)=a_i e^{H_0 t} e^{-\frac{\kappa^2\rho_0}{12}(t(t-2t_\text{i}))}\,,
\end{equation}
where $H_0$ is constant and $t_i$, as usually, is the time at the beginning of inflation, but in principle such a solution, by setting a suitable value of $\rho_0$, may reproduce also the current acceleration predicting an end of it, since it has not been used any slow-roll approximation.

\section{Conclusions}

In this paper, we have investigated viable models for inflation by starting from observations. The Planck results for the spectral index and the tensor-to-scalar ratio determine the behaviour of the models. In particular, it is possible to reconstruct a model by starting from the implicit form of this paramters: a simple Ansatz relates them to the effective EoS parameter of inflationary universe, and the Hubble parameter with the effective energy density follow from it.
We have shown how, working in scalar field representation, the only viable models which can be derived in such a way are the well known massive inflaton theory (but we must remark that in such a case the tensor-to-scalar ratio is  larger than the Planck result) and the Einstein-frame representation of the Starobinsky-like model accounting for the $R^2$-correction to Einstein's gravity. The reconstruction technique permits also to find the viable fluid models for inflation: the inhomogeneous fluids producing inflation have been analyzed and the explicit solutions have been presented. 

In the last Section, we have considered the $f(R)$-modified theories of gravity for inflation. For a modified theory of gravity the spectral index and the tensor-to-scalar ratio must be recalcualted. Thus, a reconstruction technique can be used to find the models that reproduce values in agreement with Planck data for this parameters. We found a model with an exact solution (without making use of slow-roll approximation) for viable inflation, where a $R^2$-correction together with a cosmological constant is added to the Hilbert-Einstein action of General Relativity. This model belongs to the class of Starobinsky-like inflationary theories. It is possible to see that by requiring that accelerated solution appears at high curvature, all the Starobinsky-like models can be found. As it was expected, the related spectral index and the tensor-to-scalar ratio return be the same of scalar inflation in Einstein frame representation of Starobinsky model.

\section{Acknowledgments}

S.Z. thanks G. Cognola, M. Rinaldi, and L. Vanzo for discussions.



\begin{thebibliography}{}

\bibitem{WMAP} 
G.~Hinshaw {\it et al.}  [WMAP Collaboration],
    Astrophys.\ J.\ Suppl.\  {\bf 208} (2013) 19
    [arXiv:1212.5226 [astro-ph.CO]].

\bibitem{Planck} 
 P.~A.~R.~Ade {\it et al.}  [Planck Collaboration],
  Astron.\ Astrophys.\  {\bf 571}, A22 (2014)
  [arXiv:1303.5082 [astro-ph.CO]].


\bibitem{Mukbook}
V.~Mukhanov,
    ``Physical foundations of cosmology,''
    Cambridge, UK: Univ. Pr. (2005) 421 p.

\bibitem{rev1}
A.~D.~Linde,
   Lect.\ Notes Phys.\  {\bf 738}, 1 (2008)
   [arXiv:0705.0164 [hep-th]].

\bibitem{rev2}
D.S.Gorbunov and V.A.Rubakov,
\textit{Introduction to the Theory of the Early Universe: Hot Big Bang Theory} 
(2011).



\bibitem{Guth}
A. H. Guth, Phys. Rev. D {\bf 23}, 347 (1981).

\bibitem{Sato}
K. Sato, Mon. Not. R. Astron. Soc. {\bf 195}, 467 (1981);  Phys. Lett. {\bf 
99B}, 66 (1981).


\bibitem{chaotic}
A. Linde, Phys. Lett. {\bf 129B}, 177 (1983).

\bibitem{buca1}
A. Linde, Phys. Lett. {\bf 108B}, 389 (1982).

\bibitem{buca2}
A. Albrecht and P. Steinhardt, Phys. Rev. Lett. {\bf 48}, 1220 (1982).


\bibitem{fluidOd} 
 Capozziello, S.; Cardone, V.F.; Elizalde, E.; Nojiri, S.; Odintsov, S.D. 
\emph{Phys. Rev. D} \textbf{2006}, \emph{73}, 043512:1--043512:16.

\bibitem{fluidOd2}
 S.~Nojiri and S.~D.~Odintsov,
  Phys.\ Rev.\ D {\bf 72} (2005) 023003
  [hep-th/0505215].

\bibitem{fluidOd3}
 S.~Nojiri and S.~D.~Odintsov,
  Phys.\ Lett.\ B {\bf 639} (2006) 144
  [hep-th/0606025].

\bibitem{fluidOd4}
 K.~Bamba, S.~Capozziello, S.~'i.~Nojiri and S.~D.~Odintsov,
  Astrophys.\ Space Sci.\  {\bf 342}, 155 (2012)
  [arXiv:1205.3421 [gr-qc]].




\bibitem{Alessia} 
I.~H.~Brevik and O.~Gorbunova,
Gen.Rel.Grav. {\bf 37} 2039-2045 (2005). 

\bibitem{Brevik} 
I.~H.~Brevik and O.~Gorbunova,
  Eur.\ Phys.\ J.\ C {\bf 56}, 425 (2008)
  [arXiv:0806.1399 [gr-qc]].

\bibitem{Alessia(2)} 
I.~H.~Brevik, O.~Gorbunova and Y.A. Shaido, 
Int.\ J.\ Mod.\ Phys.\  D {\bf 14} 1899 (2005).



\bibitem{mioinfl1}
 R.~Myrzakulov and L.~Sebastiani,
  arXiv:1410.3573 [gr-qc].

\bibitem{mioinfl2}
R.~Myrzakulov and L.~Sebastiani,
  arXiv:1411.0422 [gr-qc].




\bibitem{q1}
 M.~Rinaldi, G.~Cognola, L.~Vanzo and S.~Zerbini,
  arXiv:1410.0631 [gr-qc];
M.~Rinaldi, G.~Cognola, L.~Vanzo and S.~Zerbini,
  JCAP {\bf 1408}, 015 (2014)
  [arXiv:1406.1096 [gr-qc]].


\bibitem{q1bis}
 K.~Bamba, S.~Nojiri, S.~D.~Odintsov and D.~Saez-Gomez,
  Phys.\ Rev.\ D {\bf 90}, 124061 (2014)
  [arXiv:1410.3993 [hep-th]];
K.~Bamba, G.~Cognola, S.~D.~Odintsov and S.~Zerbini,
  Phys.\ Rev.\ D {\bf 90}, 023525 (2014)
  [arXiv:1404.4311 [gr-qc]];
 J.~Amoros, J.~de Haro and S.~D.~Odintsov,
  Phys.\ Rev.\ D {\bf 89}, no. 10, 104010 (2014)
  [arXiv:1402.3071 [gr-qc]].

\bibitem{q3}
K.~Bamba, R.~Myrzakulov, S.~D.~Odintsov and L.~Sebastiani,
  Phys.\ Rev.\ D {\bf 90}, 043505 (2014)
  [arXiv:1403.6649 [hep-th]];
R.~Myrzakulov, S.~Odintsov and L.~Sebastiani,
  arXiv:1412.1073 [gr-qc].


\bibitem{mioultimo}
S.~Myrzakul, R.~Myrzakulov and L.~Sebastiani,
  arXiv:1501.01796 [gr-qc].



\bibitem{mod1}
S.~Nojiri and S.~D.~Odintsov,
  eConf C {\bf 0602061}, 06 (2006)
  [Int.\ J.\ Geom.\ Meth.\ Mod.\ Phys.\  {\bf 4}, 115 (2007)]
  [hep-th/0601213];
 S.~Nojiri and S.~D.~Odintsov,
   Phys.\ Rept.\  {\bf 505}, 59 (2011)
   [arXiv:1011.0544 [gr-qc]].

\bibitem{mod2}
S.~Capozziello and M.~De Laurentis,
  Phys.\ Rept.\  {\bf 509}, 167 (2011)
  [arXiv:1108.6266 [gr-qc]].

\bibitem{mod3}
R.~Myrzakulov, L.~Sebastiani and S.~Zerbini,
  Int.\ J.\ Mod.\ Phys.\ D {\bf 22}, 1330017 (2013)
  [arXiv:1302.4646 [gr-qc]].


\bibitem{muk1}
V.~Mukhanov,
  Eur.\ Phys.\ J.\ C {\bf 73}, 2486 (2013)
  [arXiv:1303.3925 [astro-ph.CO]].

\bibitem{muk2}
V.~Mukhanov,
  arXiv:1409.2335 [astro-ph.CO].





\bibitem{C1}
   B.~Pourhassan and E.~O.~Kahya,
   Adv.\ High Energy Phys.\  {\bf 2014}, 231452 (2014)
   [arXiv:1405.0667 [gr-qc]].



\bibitem{C2}
   B.~Pourhassan and E.~O.~Kahya,
   Res.\ Phys.\  {\bf 4}, 101 (2014).






\bibitem{Staro}
   A.~A.~Starobinsky,
   Phys.\ Lett.\ B {\bf 91}, 99 (1980).

\bibitem{mioStaro}
L.~Sebastiani, G.~Cognola, R.~Myrzakulov, S.~D.~Odintsov and S.~Zerbini,
  Phys.\ Rev.\ D {\bf 89}, no. 2, 023518 (2014)
  [arXiv:1311.0744 [gr-qc]].




\bibitem{corea}
Hwang, J.-C., and Noh, H., Phys. Lett. B, 506, 13--19, (2001);
 H.~Noh and J.~c.~Hwang,
  Phys.\ Lett.\ B {\bf 515}, 231 (2001)
  [astro-ph/0107069].


\bibitem{DeFelice}
 A.~De Felice and S.~Tsujikawa,
  Living Rev.\ Rel.\  {\bf 13}, 3 (2010)
  [arXiv:1002.4928 [gr-qc]].





\end{thebibliography}
\end{document}